\documentclass[a4paper, amsfonts, amssymb, amsmath, reprint, showkeys, nofootinbib, twoside]{revtex4-1}
\usepackage[english]{babel}
\usepackage[utf8]{inputenc}

\usepackage{amsmath}
\usepackage{braket}

\usepackage{appendix}

\usepackage{graphicx}
\usepackage{hyperref}
\usepackage{tikz}
\usetikzlibrary{calc}

\usepackage{comment}

\begin{document}

\title{Many Body Density of States of a system of non interacting spinless fermions}

\author{R\'{e}mi Lef\`{e}vre}
\author{Krissia Zawadzki}
\author{Gr\'{e}goire Ithier}
\affiliation{Department of Physics, Royal Holloway University of London}

\email[Correspondence email address: ]
{gregoire.ithier@rhul.ac.uk}
\vspace{10pt}

\def\fnum@figure{\figurename\thefigure}
\renewcommand{\figurename}{Fig.}
\newcommand{\D}{\textrm{d}}% should be \mathrm{d}
\newcommand{\sech}{\textrm{sech}}% I suspect this should be \DeclareMathOperator\sech{sech}

\begin{abstract}
The modeling of out-of-equilibrium many-body systems requires to go beyond the low-energy physics and local densities of states. Many-body localization, presence or lack of thermalization and quantum chaos are examples of phenomena in which states at different energy scales, including the highly excited ones, contribute to the dynamics and therefore affect the system's properties. Quantifying these contributions requires the many-body density of states (MBDoS), a function whose calculation becomes challenging even for non-interacting identical quantum particles due to the difficulty in enumerating states while enforcing the exchange symmetry.
In the present work, we introduce a new approach to evaluate the MBDoS in the case of systems that can be mapped into free fermions. The starting point of our method is the principal component analysis of the filling matrix $F$ describing how $N$ fermions can be configured into $L$ single-particle energy levels. We show that the many body spectrum can be expanded as a weighted sum of spectra given by the principal components of the filling matrix. The weighting coefficients only involve renormalized energies obtained from the single body spectrum.
We illustrate our method in two classes of problems that are mapped into spinless fermions: (i) non-interacting electrons in a homogeneous tight-binding model in 1D and 2D, and (ii) interacting spins in a chain under a transverse field. 
\end{abstract}

\maketitle

\section{Introduction}

The concept of Density of States (DoS) 
%permeates various sub-fields of physics, in which of its importance emerges in different flavors.
%It
is at the heart of statistical physics where it 
%emerges from the definition of the 
defines partition function and temperature. In nuclear physics, quantifying the level density is necessary to describe nuclear reactions involving excited states \cite{ZELEVINSKY_2019}.   
No less importantly, it is one of the most appealing quantity in %solid state and
condensed matter physics, where one is interested in investigating how electrons and holes populate energy bands to give rise to material's properties \cite{Ashcroft1976solid}.  
%From macro, meso and microscopic scales, 
In all these fields, the density of states is crucial to characterize a multi-particle system and determine which states are accessible at energy scales of interest.

For a long time, the success of mean field theories and the quasi-particle picture to describe a highly degenerate Fermi liquid has promoted the Single-Body 
Density of States (SBDoS) and its local counterpart to the focus of investigations of electronic systems. In the presence of interactions, efforts have been concentrated in the physics at low temperatures, so that the SBDoS around the Fermi level suffices to obtain most of their properties.  Nonetheless, the quest of calculating a \textit{Many}-Body Density of States (MBDoS) has became arguably necessary
 in the context of isolated quantum systems undergoing an out-of-equilibrium dynamics \cite{Gogolin_2016}. There, contributions from different parts of the spectrum prevents one from relying on a description based solely on the low-lying states and on the SBDoS. Quantifying these contributions is crucial to shed light on phenomena such as quantum chaos \cite{Lea-PhysRevE.81.036206}, 
 thermalization and its lack \cite{BORGONOVI20161,RevModPhys.83.863,annurev-conmatphys-031214-014726},  many-body localization \cite{Bloch-RevModPhys.91.021001} and more generally unconventional stationary states 
 \cite{Ithier-PhysRevA.96.012108, Ithier-PhysRevE.96.060102}.

 In this respect, a MBDoS provides useful information, as it allows for quantifying how interactions between individual constituents lead to a complex many-body dynamics, with coexisting single-particle and collective effects. Even non-interacting systems pose a challenge due to the combinatorial nature of the problem of how single-body levels can be populated to define a distribution of many-body energy levels. For both non-interacting and interacting systems, it is possible to retrieve numerically the full spectrum of energies and eigenstates only for system sizes that do not exceed a dozen of particles. Symmetries can aid this computation by allowing one to split the total Hilbert space into blocks, associated 
 %with the projection of the Hamiltonian into states
 with conserved quantum numbers. This idea has served as the basis of exact diagonalization \cite{EDweisse2008exact} and it is also implemented in well established numerical methods, as for instance the Kernel Polynomial Method (KPM) \cite{KPM-RevModPhys.78.275}. On the analytical side,  methods to calculate the MBDoS started in nuclear physics with a Fermi gas approximation calculation derived by Bethe  \cite{Bethe-PhysRev.50.332}, which inspired approaches such as the constant temperature  or the continuum shell model (see e.g. the reviews in \cite{ZELEVINSKY_2019,Volya-PhysRevC.74.064314}). More involved methods using exact combinatorial counting
 \cite{hillman1969shell,berger1974shell}, recursive relations \cite{jacquemin1986exact}, or 
 saddle approximations \cite{bohr1998nuclear}
 %or kernel polynomial methods \cite{silver1996kernel}
 provided some approximate results.

The calculation of the MBDoS of systems of non interacting spinless fermions is the problem at the focus of the present paper. We propose a new approach to calculate the exact MBDoS based on the symmetries of a rectangular filling matrix describing how $N$ particles can be combined into $L$ single-particle energy levels to generate the many-body states. The starting point of our method relies on the singular value decomposition of this matrix, which allows to expand the many-body spectrum as a weighted sum of 'principal' spectra. These spectra depend \textit{only} on the number of single body levels and the number of particles. The weighting factors involve a discrete Fourier transformation of the single body energies, providing renormalized energies. 
%We show that it is possible to reconstruct the right eigenvectors associated with the largest dimension of the filling matrix by exploring its symmetries. 
Our approach is illustrated in the case of spinless non-interacting fermions, and applied to the tight-binding model in one and two dimensions, and to the transverse Ising field chain.

We organize the paper as follows. In Sec. \ref{sec:PCA}, we introduce the idea of principal analysis decomposition of a filling matrix and explain how it allows to access a non interacting Many Body spectrum. In Sec. \ref{sec:SpinlessFermions}, we develop the method in the case of spinless fermions, discuss how to explore symmetries of the problem of calculating the MBDoS. Applications to tight-binding and Ising chains are discussed in Sec. \ref{sec:Applications}. Finally, our main findings are summarized in Sec.  \ref{sec:Conclusions}.

\section{Principal Components Approach}
\label{sec:PCA}

We start by considering a system for which we know the single-body spectrum given by the energies $\epsilon_k$. Our goal is to construct the many-body
spectrum, which has energies $E_p =\sum_{k=1}^L  \epsilon_k f_p^k$, where $f_p^k$ is the filling factor or occupation
number of the $k^{th}$ single body level for the $p$-th many-body state. %i.e. a configuration of particles dispatched over the $L$ single body levels. 
Our first step is to rewrite all those energies in
matrix form by collecting all single-body energies $\epsilon_k$ into a column vector $\epsilon$, all many body energies $E_p$ into a column vector $E$, and finally construct
a rectangular matrix $F$, the "filling" matrix, of general term $f_p^k$.
We can then write the relation between many body and single body energies in matrix form :
\begin{equation}
    \label{mb_energy_vector}
    E = F \cdot \epsilon
\end{equation}
where $\cdot$ denotes the matricial product.

The singular value decomposition (SVD) of the filling matrix $F = U \cdot \Sigma \cdot V^\dagger$ with $U,V$ square unitary matrices and $\Sigma$ rectangular diagonal,
provides its so called ''principal component''
expansion :
\begin{equation}
    \label{mb_filling_matrix}
    F = \sum_{\ell=0}^{L-1} \sigma_\ell \; U_\ell \cdot V_\ell^\dagger
\end{equation}
where $U_\ell$ is the $l^{th}$  right-singular vector (i.e. the $l^{th}$ column of $U$),
$\sigma_\ell$ are the singular values (i.e. the main diagonal of $\Sigma$), and
$V_\ell$ is the $l^{th}$ left-singular vector (i.e. the $l^{th}$ column of $V$). 
%Note that $\tilde{u}_\ell \tilde{v}_\ell^\dagger$ is the outer product of the two vectors $\tilde{u}_\ell, \tilde{v}_\ell$. 
To ease notations, in the following 
%we will work with right-singular vectors which are orthonormal, denoted by $V_\ell = \tilde{v_\ell}$ but with left-singular vectors which are not normalized such that 
we will incorporate the singular value $\sigma_l$ into the definition of $U_l$ so that the 
right-singular vectors are orthogonal but \textit{not} normalized anymore.
%$u_\ell = \sigma_\ell \tilde{u_\ell}$. 
Note that with these
notations, $U_\ell$ and $V_\ell$ are simply related by $U_\ell = F \cdot V_\ell$. By combining Eqs. \ref{mb_energy_vector} and \ref{mb_filling_matrix}, the vector $E$ of many-body energies can be rewritten as :
\begin{equation}
    \label{mb_spectral_expansion}
    E = \sum_{\ell=0}^{L-1} U_\ell (V_\ell^\dagger \cdot \epsilon)
        = \sum_{\ell=0}^{L-1} \tilde{\epsilon}_\ell \; U_\ell
\end{equation}
where we defined the scalars $\tilde{\epsilon_\ell} = V_\ell^\dagger \cdot \epsilon$.

The many body spectrum now appears as a sum of "principal" \textit{spectral} components
given by the $U_\ell$ vectors. These spectral components are weighted by the new effective
energies $\tilde{\epsilon_\ell}$ of a \textit{renormalized} single-body spectrum. In particular,
the form of $V_\ell$ defines which of these energies contribute to the many-body spectrum.
As we will discuss later, a very convenient choice is given by Fourier modes obtained from
the analytical solution for the eigenvectors of the circulant matrix $F^\dagger \cdot F$. Depending on
the band structure of the system, it is also possible that some renormalized energies
vanish which allows the associated $\ell$ modes to be discarded.

The problem of computing the many body spectrum is now split in two parts. The first part
is to compute the renormalized single body spectrum only depending on the right singular
part of the SVD (ie. the $V_\ell$) which can be obtained analytically or for a small computational cost scaling like $L$. 
The second part depends on the left singular part of the SVD (ie. the $U_\ell$ incorporating the $\sigma_l$)
%which, in our notation, incorporates the singular values,
which only contain information about the universal properties of many-body
systems encoded in the combinatoric structure of the $F$ matrix. That second part requires
a statistical approach but can be efficiently computed for large system sizes by avoiding
using the $F$ matrix explicitly since its number of rows scales exponentially with the number of
levels and particles considered.
It should be noted that despite the $F$ matrix is obviously a real matrix, the renormalized energies and the $U_\ell$ vectors can be complex, as we will see in the following.

In the following, we provide a framework to compute the singular decomposition of $F$
in the case of a system of non-interacting fermions with no other quantum number.

\section{Spinless Fermions}
\label{sec:SpinlessFermions}

In the case of fermions without any additional quantum number, occupation numbers can only
be $0$ or $1$, which are the possible matrix components of $F$. Each row of the filling
matrix $F$ (i.e. some configuration of a many body state) is a binary string. All rows
share the same amount of $1$'s, to account for the fixed number of particles $N$, there are
a total of $C_L^N$ many body state configurations and the filling matrix $F$ has dimensions
$L \times C_L^N$.

\subsection{Right part of the SVD}

We focus first on the right-singular vectors $V_\ell$ and the singular values
$\sigma_\ell$. Both can be computed by considering the eigendecomposition of the square matrix
$F^\dagger \cdot F$ which has dimensions $(L \times L)$.
Each matrix element $(F^\dagger \cdot F)_{i,j}$ is the scalar product between the $i^{th}$ and $j^{th}$ columns of $F$,  each encoding the filling factor of the $i^{th}$ and $j^{th}$ single body states along all of the many body states. If $i=j$, simple combinatorics considerations indicate there are
$C_{L-1}^{N-1}$ many body levels with the $i^{th}$ single body level filled with one particle.
If $i \neq j$, there are 
$C_{L-2}^{N-2}$ many body levels with both the $i^{th}$ and $j^{th}$ single body levels filled with a particle. From this reasoning we deduce that the $F^\dagger \cdot F$ matrix has the form of a circulant matrix with only two possible values :
\begin{equation*}
    F^\dagger \cdot F = \begin{pmatrix}
        a & b & \dots & b \\
        b & \ddots & & \vdots \\
        \vdots & & \ddots & b \\
        b & \dots & b & a
    \end{pmatrix}_{L \times L}
\end{equation*}
where $a = C_{L-1}^{N-1}$ and $b = C_{L-2}^{N-2}$. 
Circulant matrices
are well known and can be diagonalized using Fourier modes.
Let $\omega = e^{i 2\pi / L}$, then eigenvectors of $F^\dagger \cdot F$, which are the right-singular
vectors $V_\ell$ of $F$, are given by :
\begin{equation}
    \label{mb_svd_ul}
    V_\ell = \frac{1}{\sqrt{L}}
        \left( 1, \omega^{\ell}, \omega^{2\ell}, \dots, \omega^{(L-1)\ell} \right).
\end{equation}
The eigenvalues of $F^\dagger \cdot F$ take the following form :
\begin{equation*}
    \lambda_\ell = C_{L-1}^{N-1} + C_{L-2}^{N-2} \sum_{k=1}^{L-1} \omega^{k\ell}
\end{equation*}
from which only two distinct values can arise. After taking the
square root to obtain singular values of $F$, we get :
\begin{align}
    \label{mb_svd_eigenvalues}
    \sigma_0 &= \sqrt{N C_{L-1}^{N-1}} &\text{ with multiplicity } 1 \nonumber \\
    \sigma_{\ell \ge 1} &= \sqrt{C_{L-2}^{N-1}} &\text{ with multiplicity } L-1
\end{align}

One can easily check that the $V_\ell$ vectors form an orthonormal base associated to these eigenspaces,
the $\ell=0$ case matching the $1$-dimensional eigenspace spanned by $V_0$ and the
$\ell \ge 1$ case matching the $L-1$-dimensional eigenspace spanned by $V_{\ell \ge 1}$.

\subsection{Left part of the SVD}

In order to compute the left-singular vectors $U_\ell$, we avoid dealing with the matrix
$F \cdot F^\dagger$ which is much larger than $F^\dagger \cdot F$, and prefer using the identity $U_\ell = F \cdot V_\ell$. We can
write the components of $U_\ell$ as follows :
\begin{equation*}
    U_\ell^p = \frac{1}{\sqrt{L}}\sum_{k=0}^{L-1} F_{k}^p \omega^{k\ell}
\end{equation*}
essentially, the inner product between $V_\ell$ and the configuration of the $p^{\text{th}}$
many body state.

We will now investigate two symmetries which provide a good understanding of universal
properties of $F$ and allow to reduce the very large set of configurations to a more
manageable size for numerical applications.

\subsection{The $k$-symmetry}

\begin{figure}[htb]
    \centering
    \includegraphics[width=\columnwidth]{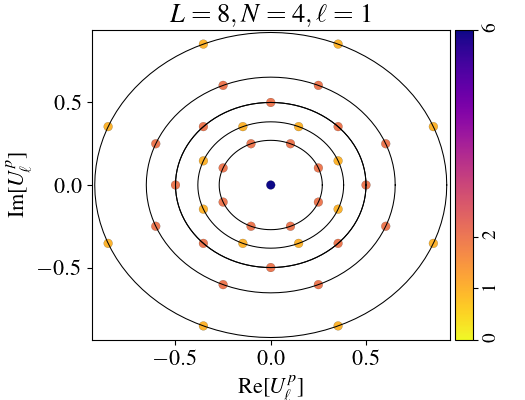}
    \caption{
        \textbf{First principal spectrum $U_1$ of the filling matrix $F$ for $L=8$ single body levels and $N=4$ spinless fermions.}
        Each point is a distinct value of the distribution of the components of $U_1$ in the complex plane.
        Circles relate components belonging to one or more equivalence classes having the same
        modulus. The number of occurences of each values are : $6$ (dark purple) for all
        degenerate classes with radius $0$, $1$ (yellow) which have $1$ non-degenerate class
        per circle, and $2$ (dark orange) which have $2$ non-degenerate classes per circle.
    }
    \label{fig:circles_Vls}
\end{figure}

The first symmetry we observe is on the $V_\ell$ vectors containing Fourier modes based on
$L^{\text{th}}$ roots of unity. Since they contain components of the form $\omega^{k\ell}$,
when calculating the inner product of $V_\ell$ against a configuration
represented by a binary string, a circular permutation of those bits will multiply the
Fourier modes by a power of $\omega$, i.e. it will rotate the result $U_\ell^p$ in the
complex plane by a complex factor which is a power of $\omega^\ell$. We will denote the
associated angle $\theta_{\text{step}} = 2 \pi \ell / L$. Note that it does not depend on
$N$ and remains valid for all components of a given $U_\ell$.
It follows that all the components of $U_\ell$ lie on various circles in the complex
plane, and are located at relative angles which are multiples of $\theta_{\text{step}}$.
We call this structure the $k$-symmetry, which is illustrated on Fig. \ref{fig:circles_Vls}.

The $k$-symmetry suggests to group configurations together in equivalence classes defined
by the underlying relation of circular permutations. Each class contains configurations that
can be transformed into each other by circular permutation, we choose the lowest of them in
lexicographic order as class representatives which we will refer to as "seeds". All members
of such an equivalence class, when taking the inner product with $V_\ell$, will produce
$U_\ell$ components which lie on the same circle in the complex plane, for any fixed given
$\ell$. Note that two distinct classes can still map to the same circle by having the same
radius.

There will be two kind of such equivalence classes : the non-degenerate classes contain
exactly $L$ binary strings when the bits don't provide any additional symmetry, while the
degenerate classes contain less than $L$ strings as the bits expose a shorter pattern that
repeats under circular permutation, for instance $(0,0,1,1,0,0,1,1)$ is an $8$-bits sequence
but loops back early with a circular permutation by $4$ positions. This distinction will
become important later on.

Next, we notice that equivalence classes are closely related to the factors of $L$. In
particular, if $L$ is a prime number, all classes are non-degenerate. Degenerate classes
can only have a cardinal which has a common factor with $L$. This can be deduced from
the action of circular permutations on strings of various sizes.
This is of particular physical meaning : the prime decomposition of $L$ is a decisive
feature for the symmetries of the filling matrix and its singular value decomposition.

Finally, a counting argument can be made about those equivalence classes by using the
Pólya enumeration theorem, from which we can estimate the number of equivalence classes
to be of the order of $C_L^N / L$.

\subsection{The $\ell$-symmetry}

\begin{figure*}
    \centering

    \begin{tikzpicture}
        \node[label={[font=\normalsize, shift={(0cm,0mm), align=center}]above:(a)}] (figa) at (0,0)
        {\includegraphics[ trim={0cm 0.cm 0cm 0.cm}, clip, width=2\columnwidth]{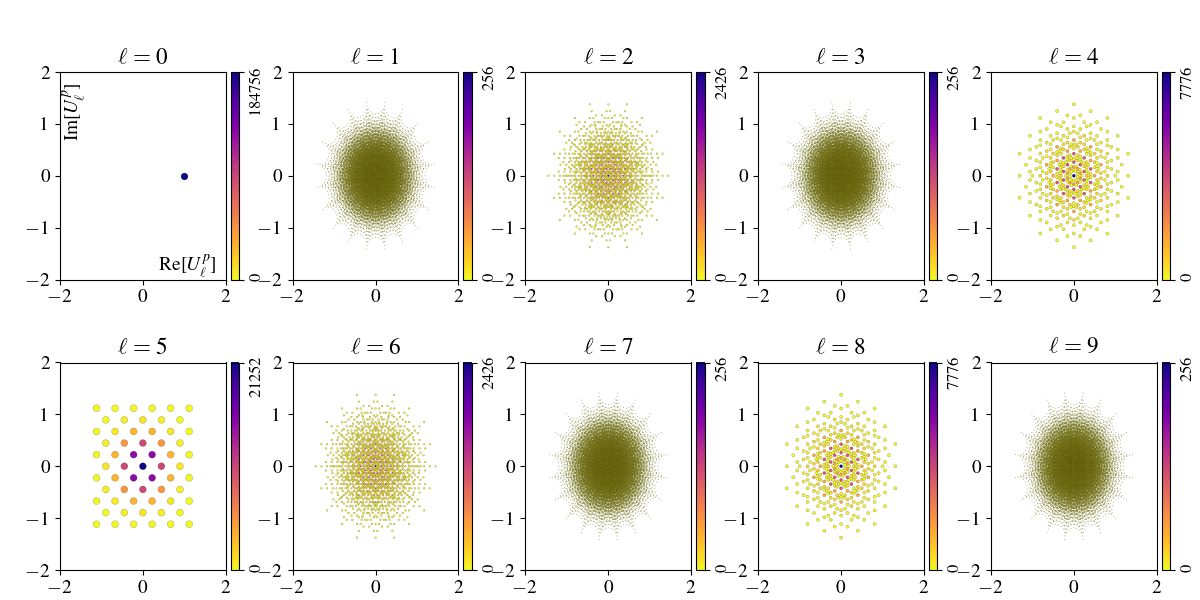}};
        
        \node[anchor=north west, 
       label={[font=\normalsize, shift={(0cm,0mm), align=center}]above:(b)}
        ] (figb) at ($(figa.south west)+(-0.0cm,-0.75cm)$)
        {\includegraphics[ trim={0cm 0.cm 0cm 0.cm}, clip,width=2\columnwidth]{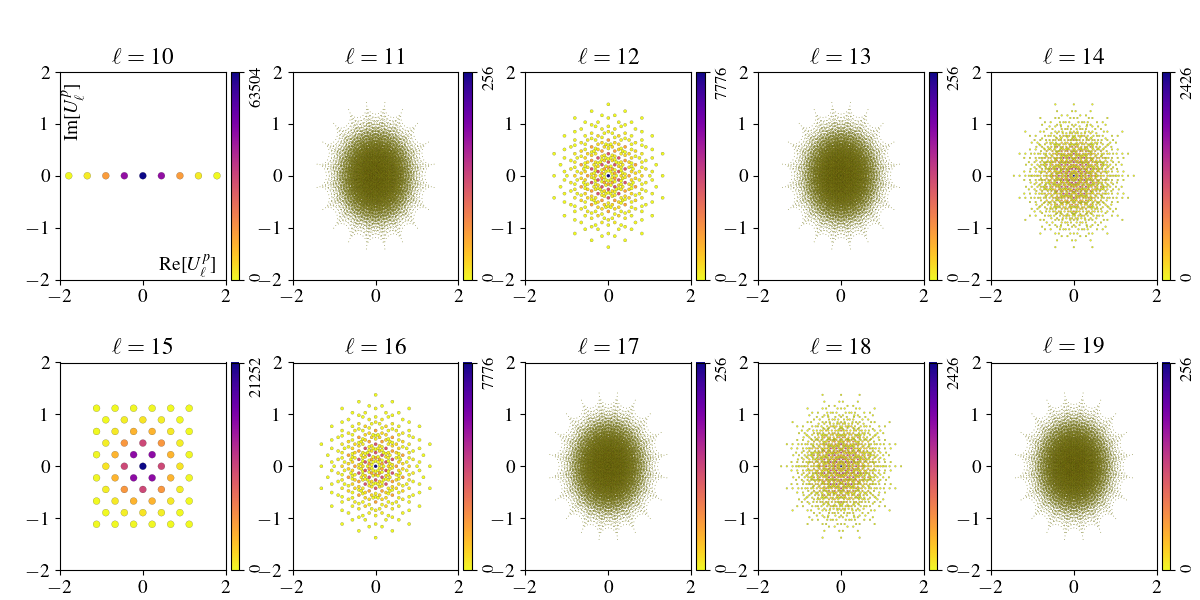}};
    \end{tikzpicture}

    \caption{
        \textbf{Full spectrum of components of $U_\ell$'s for $L=20$ at half-filling $N=10$.}
        All $\ell$ values with the same greatest common divisor with $L$ are associated to the same principal spectrum $V_\ell$, illustrating the $\ell$-symmetry discussed in
        Sec.\ref{sec:SpinlessFermions}.
    }
    \label{fig:full_spectrum_Vl}
\end{figure*}

On the other hand, we can look at what happens when we consider a different Fourier
mode and go from some $\ell$ to another $\ell'$. 
As $\omega$ is by definition a \textit{primitive} $L^{th}$ root of unity (i.e. $\omega^L=1$ but $\omega^p \neq 1$ for all $p <L$), then each $L^{th}$ root of unity is a distinct power of $\omega$. It is well known that $\omega^\ell$ is a $r^{th}$ primitive root of unity for
$r=L/\text{gcd}(L,\ell)$. Considering another value $\ell'$, if
\begin{equation}
    \label{mb_lsymmetry_criterion}
    \gcd(L,\ell) = \gcd(L,\ell')
\end{equation}
%gcd$(L,\ell')=$ gcd$(L,\ell)$ 
then $\omega^{\ell'}$ is also a 
$r^{th}$ primitive root of unity. In other words, each power of $\omega^\ell$ is in a one to one correspondance with a power of $\omega^{\ell'}$, so the components of $V_{\ell'}$ are just a permutation of the components of $V_\ell$.
In this case, since the permutation of $V_\ell$ components is equivalent to the permutation
of configurations in the $F$ matrix, then vectors $U_\ell$ and $U_{\ell'}$ are also the
same up to a permutation, i.e. the distributions of their components are identical and the resulting spectra are identical. We call this property the $\ell$-symmetry, verified when \eqref{mb_lsymmetry_criterion} is valid, which is illustrated in
Fig.\ref{fig:full_spectrum_Vl} where all principal spectra are provided for $L=20$ and $N=10$. 

In order to proceed further, we need to introduce the language of compositions. A composition of an
integer is similar to a partition, but with order taken into account \cite{knuth2015art}. Given some integer $n$ (in our case the number of particles),
we decompose $n$ into a sum of $p$ non-zero integer called the "parts". Any
such sequence is called a $p$-composition of $n$. If we allow some of the $p$ parts to be
zero, then it is called a weak $p$-composition of $n$. Finally, if we also impose some integer
$k$ such that any part can only have a maximum value of $k$, then such list of integers is called  a $k$-restricted
weak $p$-composition of $n$.

The $\ell$-symmetry simplifies the study of the principal spectra, as only a small amount of
$\ell$ values are required to obtain the complete set of $U_\ell$, namely the divisors of $L$.
To implement this, we can change our point of view on binary strings, and consider them as
$1$-restricted weak $L$-compositions of the integer $N$. We then proceed by considering only
the $\ell$'s which divide $L$, and introduce the integer $q = L/\ell$. We subdivide each
binary string into $\ell$ sections of size $q$, and add them together component by component,
effectively folding them into a vector of size $q$. We obtain $\ell$-restricted weak
$q$-compositions of the integer $N$. This leads to the following simplification :
\begin{equation*}
    U_\ell^p = \frac{1}{\sqrt{L}}\sum_{k=0}^{q-1} \left( \sum_{s=0}^{\ell-1} F_{sq+k}^p \right) e^{i 2\pi \frac{k}{q}}
\end{equation*}
from which we redefine effective vectors $V'_\ell$ of size $q$, which now contain Fourier
modes based on the $q^{\text{th}}$ roots of unity $\omega' = e^{i 2\pi / q}$. Its components
are of the form $\omega'^k$ with $k \in \{0,q-1\}$.

\begin{figure*}[htb!]
    \centering
    \begin{tikzpicture}
        \node[label={[font=\normalsize, shift={(0cm,0mm), align=center}]above:(a)}] (figa) at (0,0)
        {\includegraphics[ trim={0cm 0.cm 0cm 0.cm}, clip, width=2\columnwidth]{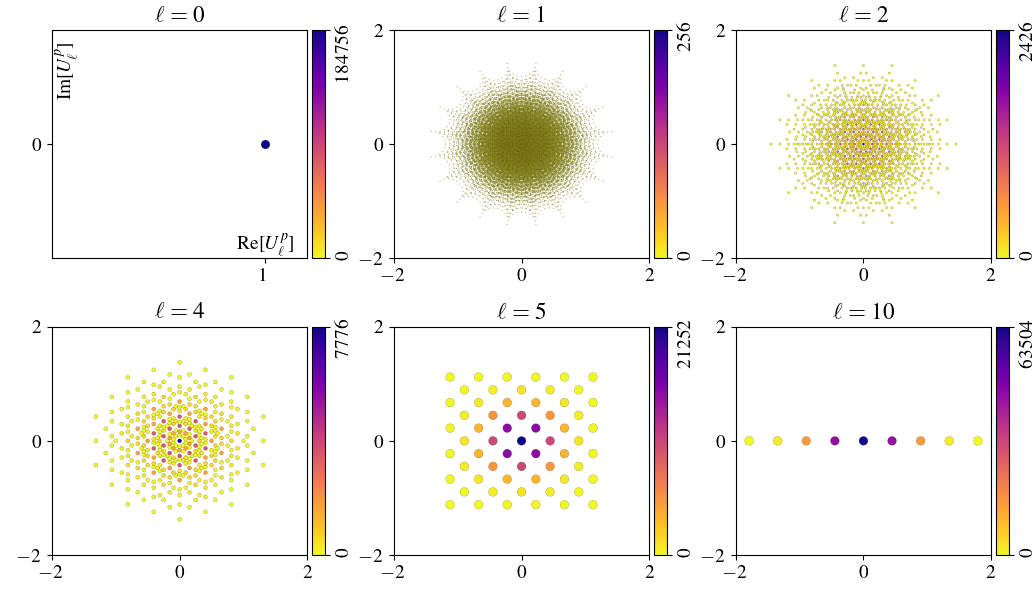}};
        \node[anchor=north west, 
       label={[font=\normalsize, shift={(0cm,0mm), align=center}]above:(b)}
        ] (figb) at ($(figa.south west)+(-0.0cm,-0.75cm)$)
        {\includegraphics[ trim={0cm 0.cm 0cm 0.cm}, clip,width=2\columnwidth]{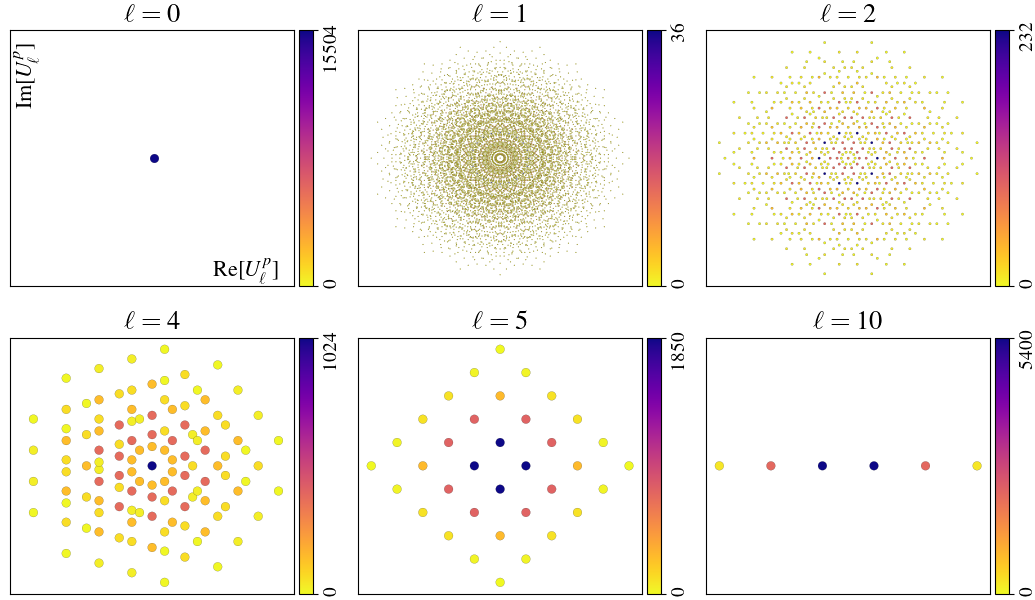}};
    \end{tikzpicture}

    \caption{
        \textbf{Principal spectra of the filling matrix $F$ in the case $L=20$ at half-filling
        $N=10$ (a) or quarter-filling $N=5$ (b).}
        Each graph represents components of the $U_\ell$ vector in the complex plane and their occurrence numbers.
        All relevant $\ell$ values with distinct greatest common divisors with $L$ are shown,
        other $\ell$ values yield identical distributions through $\ell$-symmetry, according
        to the criteria from Eq.\eqref{mb_lsymmetry_criterion}.
        The full set of $\ell$ values can be found in Fig. \ref{fig:full_spectrum_Vl}.
        The $\ell = 0$ case is also shown and corresponds to the $1$-dimensional eigenspace
        associated to the singular decomposition of $F$.
        One can notice that increasing values of $\ell$ yield more clustered distributions.
    }
    \label{fig:unique_spectrum_Vl}
\end{figure*}

Once again, we see that the divisors of $L$, i.e. its prime decomposition, plays a central
role. In particular, if $L$ is prime, we only need to compute the $\ell=1$ case where the
$\ell$-symmetry is trivial and the $\ell$-restricted weak $q$-compositions are the binary
strings themselves. For a composite $L$, there is an interesting feature which can be observed from
Fig.\ref{fig:unique_spectrum_Vl}, where increasing $\ell$ values lead to a fast decreasing
density of $U_\ell$ components in the complex plane, while the occurence counts of each points
necessarily goes up : we transition from a spread distribution to a clustered distribution,
i.e. with larger degeneracies, and a lattice pattern can emerge. The full understanding of the rich and qualitatively different behaviors displayed on Fig.\ref{fig:unique_spectrum_Vl} as $l$ increases  requires further study and will be the subject of another article.

\subsection{Enumeration and statistics}

Both symmetries described above simplify the study of the problem. Going from the set of
$C_L^N$ binary strings to a set of $\ell$-restricted weak $q$-compositions is not an
injection, i.e. distinct binary strings can have the same weak composition, so occurrence
numbers should be tracked.
Let's denote by $\{m_0,\dots,m_{q-1}\}$ the parts of an $\ell$-restricted weak $q$-composition,
with $m_k \in \{0,\ell\}$ and their sum adding up to $N$. Then, the total number of binary
strings mapped to the same composition is given by a winding factor :
\begin{equation}
    \label{mb_composition_winding}
    Q = \prod_{k=0}^{q-1} C_\ell^{m_k}
\end{equation}

From this point on, all "folded" configurations are now represented by a choice for each
of the $m_k$ values and we use $V'_\ell$ vectors of size $q$ containing $q$-roots of
unity $\omega'^k$ for $k \in \{0,q-1\}$. For notation simplicity, we will now drop the
primes and simply redefine $V_\ell$ and $\omega$.

The next step is to apply the $k$-symmetry on top of compositions, by again noticing that
any circular permutation of a composition is only a multiplication of $V_{\ell}$ by a power
of $\omega$ : we group compositions together in equivalence classes related by circular
permutations. This is where an interesting feature comes in : by applying the $k$-symmetry
on compositions, it can be shown that configurations inside degenerate classes will always
have a null projection on the Fourier modes of $V_\ell$. We only need to count them to
obtain the occurence count of zero components in $U_\ell$.

We now have all the tools required for our framework. We can describe a simple recipe
to obtain the exact distribution of all $U_\ell$ vectors. First, we look at the value of
$L$ and construct the list of all its divisors : this gives us the $\ell$ values we have
to deal with. For each of them, we compute $q = L/\ell$ and proceed by building the list of
seeds : all $\ell$-restricted weak $q$-compositions which are not related by circular
permutation. For each seed, we first look at the size of its equivalence class. If it is
degenerate, i.e. lower than $q$, then the class lies on a $0$-radius "circle" which is a
point at origin in the complex plane. In this case, we only need to recover the occurrence
count which is the winding factor $Q$ from Eq.\eqref{mb_composition_winding} times the
degenerate class size. Otherwise, the class is non-degenerate and has size $q$ : the
total occurrence count is $Q$ times $q$, each vertex having an occurrence count of $Q$.
We also compute the inner product of the seed with $V_\ell$ directly and extract the module
and argument of the result. The module needs to be normalized by $1/\sqrt{\ell}$ to give
us the radius of this seed's circle. The argument gives us one point on the circle from
which we can reconstruct all other points coming from this seed's equivalence class by
using increments of $\theta_{\text{step}}$, i.e. the $k$-symmetry.
We now know everything about that particular seed and its class. While processing all
unique seeds and classes, we can accumulate the ones found to lie on the same circle, i.e.
having the same radius, and simply add up occurrence counts accordingly. The end result
is a list of all the circles, their respective radius, angular positions of all vertices
if needed, and occurrence numbers (by point or by circle, whichever is needed). This gives
the exact distribution of $U_\ell$ components. Finally, all other $\ell$ values which were
skipped are recovered from $\ell$-symmetry by the criteria from
Eq.\eqref{mb_lsymmetry_criterion}.

As an example, let us consider the case $L=8$ at half-filling $N=4$. There are a total of $70$ possible
many body configurations written as binary strings. The divisors of $L$ are $\{1,2,4\}$.
For $\ell=1$ (i.e. binary strings), the $k$-symmetry gives $8$ non-degenerate classes of
size $8$ and $2$ degenerate classes of size $4$ and $2$, for a total of $10$ seeds.
For $\ell=2$, applying both symmetries, we have $2$-restricted weak $4$-compositions,
$4$ non-degenerate classes of size $4$ and $2$ degenerate classes of size $2$ and $1$,
for a total of $6$ seeds.
For $\ell=4$, applying both symmetries, we have $4$-restricted weak $2$-compositions,
$2$ non-degenerate classes of size $2$ and $1$ degenerate class of size $1$, for a
total of $3$ seeds.
As one can see, the number of configurations to consider is significantly reduced. For
larger systems, e.g. $L=20$ at half-filling $N=10$, the number of many body configurations
written as binary strings is $184756$. In the worse case, $\ell=1$, this
is reduced to $9252$ binary seeds. In better cases like $\ell=4$, using $\ell$-symmetry leads
to a much smaller set of $381$ compositions which is further reduced to only $77$ seeds.

For even larger systems, ie. $L$ of the order of $100$, the lower $\ell$ values still require
significant computational time, mostly for $\ell=1$, unless they are truncated
%by only considering $\ell$ values in descending order, 
or a statistical approach is added to sample
the set of seeds. However, it should be noted that those computations are independent of
the single body spectrum: they should be performed only once for relevant values of $L$
and $N$ (ie. half-filling, quarter-filling, ...) and re-used for many different systems.
Once the distribution of $U_\ell$ components is known, the many-body spectrum and its DoS are obtained from Eq. \eqref{mb_spectral_expansion}.
The spectrum for $l=1$ which is of particular importance for the applications is displayed on  Fig.\ref{fig:MBDoS_TB1D}.

\section{Applications}
\label{sec:Applications}
% 1D, 2D and 3D single body densities of states. 
% 

% Moved from section 3 to section 4

A variety of systems can be described in terms of spinless fermions,  including hard core bosons \cite{GirardeauJMP-16.1960}, Mott insulators in ladders \cite{Donohue}, spin liquids \cite{Glazman-RevModPhys.84.1253}, and, very recently, they served as the basis to study topological phases  \cite{Turner-PhysRevB.83.075102} and systems supporting Majorana fermions \cite{Kitaev_2001, Alicea-PhysRevB.81.125318}. 

To illustrate our method, we will apply it to two classes of problems:  (i) tight-binding describing electrons  in a one-dimensional chain and in a square lattice, and (ii) the transverse Ising field chain that describes a critical 1D spin chain, and which can be mapped into a single-particle problem using the Jordan Wigner (JW) transformation. In both, we consider periodic boundary conditions (PBC).

% maybe the Hubbard model in infinite dimension because the DoS is Gaussian and the energies as well ?????

In the case (i), the Hamiltonian reads
\begin{align}
    \label{eq:tight-binding}
    H = -t \sum_{\braket{i,j}} ( c_{j}^\dagger c_{j} + \text{H.c.}),
\end{align}
where the sum $\braket{i,j}$ runs over first-neighbor sites, $c_i/c_i^\dagger$ are the annihilation/creation operators, and $t$ is the hopping amplitude.
In 1D, the single-body energies are
\begin{align}
    \label{eq:ek_TB}
    \epsilon_{k} & = -2t \cos(k),
\end{align}
the momenta $k$ depends on the boundary conditions of the model. For periodic boundary conditions (PBC) $k = \frac{2\pi n}{L}$, with $n = 0, ..., L-1$.

\begin{figure}
    \centering
    \begin{tikzpicture}
        \node[label={[font=\normalsize, shift={(3mm,-5mm), align=center}]above:(a)}] (figa) at (0,0)
        {\includegraphics[ trim={0cm 0.cm 0cm 0.cm}, clip, width=\columnwidth]{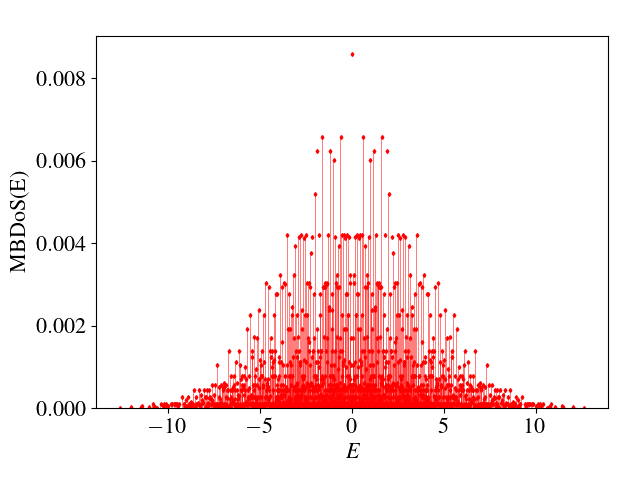}};
        
        \node[anchor=north west, 
       label={[font=\normalsize, shift={(3mm,-5mm), align=center}]above:(b)}
        ] (figb) at ($(figa.south west)+(-0.0cm,-0.75cm)$)
        {\includegraphics[ trim={0cm 0.cm 0cm 0.cm}, clip,width=\columnwidth]{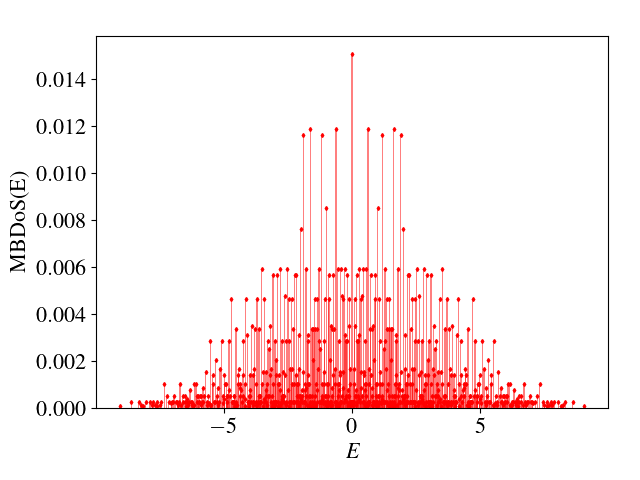}};

    \end{tikzpicture}
    \caption{
        \textbf{Many-body DoS for a 1D tight-binding chain with $L=20$ sites at half-filling
        $N=10$ (a), or quarter-filling $N=5$ (b).}
        These histograms of the many body energies are are obtained by applying Eq.\eqref{mb_spectral_expansion} where the only non-zero renormalized energies $\tilde{\epsilon}_\ell$ are for $\ell=1$ and $\ell=L-1$
        (see Sec. \ref{sec:Applications}) and 
        with $80$ bins from the lowest to the highest many body energies. The MBDoS is then normalized to a probability density.
        Note that the particle-hole symmetry explains the even parity of the profile. 
    }
    \label{fig:MBDoS_TB1D}
\end{figure}

The trigonometric form of the dispersion relation associated with the fact that the matrix Hamiltonian is circulant has an interesting implication to the renormalized energies in Eq.(\ref{mb_spectral_expansion}). 
The periodicity shared between the single-body energies and the Fourier modes allows for the cancellation of all renormalized energies $\tilde{\epsilon}_\ell$ except those associated with $\ell=1$ and $\ell=L-1$, which are symmetric to each other. 
We can show that these surviving contributions are equal to $ \tilde{\epsilon}_1 = \tilde{\epsilon}_{L-1} =  -t \sqrt{L}$. 
%This result is illustrated in Fig. \ref{fig:new_single_particle_energies}.
As a consequence, the Many Body DoS involves only one principal spectrum, the $l=1$ and is displayed on Fig. \ref{fig:MBDoS_TB1D}.

In 2D, the band structure contains $L^2$ single-body energies given by 
\begin{align}
    \label{eq:ek_TB_2D}
    \epsilon_{k_x, k_y} & = -2t \cos(k_x) - 2t \cos(k_y),
\end{align}
where the momenta in each direction are $k_x = \frac{2\pi n_x}{L}$
 and $k_y = \frac{2\pi n_y}{L}$, with $n_x, n_y = 0, ..., L-1$.
 
\begin{figure}[htb!]
    \centering
    \includegraphics[width=0.9\columnwidth]{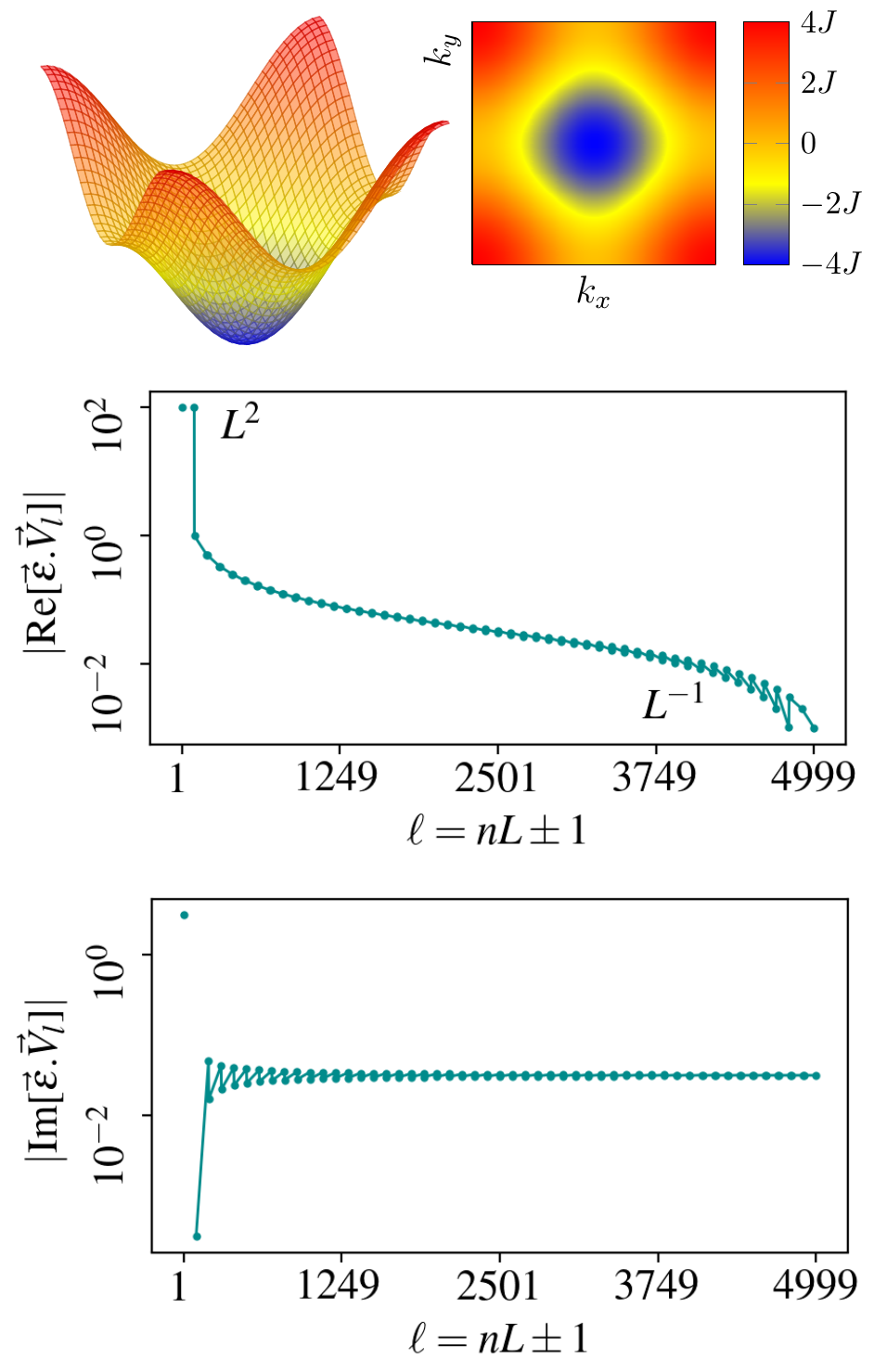}
    \caption{
        \textbf{Band structure and renormalized energies of a homogeneous square lattice with
        $L =100$ sites in each direction.}
        Top panels show the single-body band structure $\epsilon_{k_x,k_y}$ as a
        surface in 3D (left) and as a contourplot (right).
        Bottom panels display the real and imaginary part of the renormalized energies $\tilde{\epsilon}_\ell$.
        %= V_\ell^\dagger \cdot \epsilon$.
        Note that only $2(L+1)$ renormalized energies indexed as $\ell = nL\pm1$ do not vanish.
    }
    \label{fig:new_single_particle_energies_TB2D}
 \end{figure}

 Similarly to the one-dimensional case, the renormalized band structure of the square lattice will be non zero for only a few $\ell$'s, due to the high degeneracy of the single-body energies. In particular, $\epsilon_{k_x, k_y} = 0$ has degenegeracy $2(L-1)$, whereas the only non-degenegerate energies occur at the bottom and at the top of the band, where $\epsilon_{k_x, k_y} = -4J$ and  $\epsilon_{k_x, k_y} = 4J$, respectively. By flattening the 2D band structure as a 1D vector with $L^2$ entries, one can show that the $\tilde{\epsilon}_\ell$ do not vanish for $\ell = nL + 1, n = 0,1,..., L/2-1$ and $\ell = mL-1, m=1,2,...,L/2$, and for $\ell=L$, which is a special case in which $\tilde{\epsilon}_\ell$ is real. 
 The band structure in Eq. \eqref{eq:ek_TB_2D} and the non-vanishing renormalized energies $\tilde{\epsilon}_\ell$ for a square lattice with $L = 100$ sites is shown in Fig. \ref{fig:new_single_particle_energies_TB2D}. The real part decays fast with $\ell$, while the imaginary part converges to a constant, with oscillations decreasing with $\ell$.
 
In the case (ii), the Hamiltonian is 
\begin{align}
    \label{eq:TIFM}
    H & = - J \sum_j \sigma_j^x \sigma_{j+1}^x - h \sum_j \sigma_j^z,
\end{align}
where $\sigma^d$ $d=x,y,z$ are the Pauli matrices, $J$ is the coupling and $h$ is the transverse field. 

The critical point occurs at $h=J$. After employing the JW transformation, we can obtain the dispersion relation of this model as follows 
\begin{align}
    \label{eq:ek_TFIM}
    \epsilon_{k\pm} & = \pm 2J \sqrt{h^2 + 1 -2h \cos(k)} - 2Jh,
\end{align}
with $k = \frac{2n\pi }{L}$ and $n = 0, ..., L-1$ being the momentum. Note that now, two bands, one positive and the other negative, contribute to the density of states. 

In this case, the structure of the momenta $k$ corresponding to the Fourier modes result in most of the renormalized energies to be non-zero, however only odd indexes contribute. For the real part, all $\text{Re}[\tilde{\epsilon}_\ell] = 4 / \sqrt{2L} |h/J-1|$ with odd $\ell$ are independent of $\ell$. The factor in modulus introduces a symmetry around the critical point $h_c = J$, so that $\tilde{\epsilon_\ell}[h-h_c<0] = \tilde{\epsilon_\ell}[h-h_c>0]$.
For the imaginary part, the  $\text{Im}[\tilde{\epsilon}_\ell]$ %associated with $l$ indexes
are non-zero but decay very fast. The decay rate is amplified as the system size increases. 
At the critical point $h=J$, the only surviving single-body energies are
$\tilde{\epsilon}_{1} = - \tilde{\epsilon}_{L-1} = 2i \sqrt{2L}$. In other words, we recover the same renormalized single body energies as for the 1D tight binding model.

\begin{figure}[htb]
    \centering
    \begin{tikzpicture}
	\node[anchor=north west] (TFIM-contour) at (0,0) {\includegraphics[trim={2.5cm 1cm 1.75cm 2.5cm}, width=0.85\columnwidth]{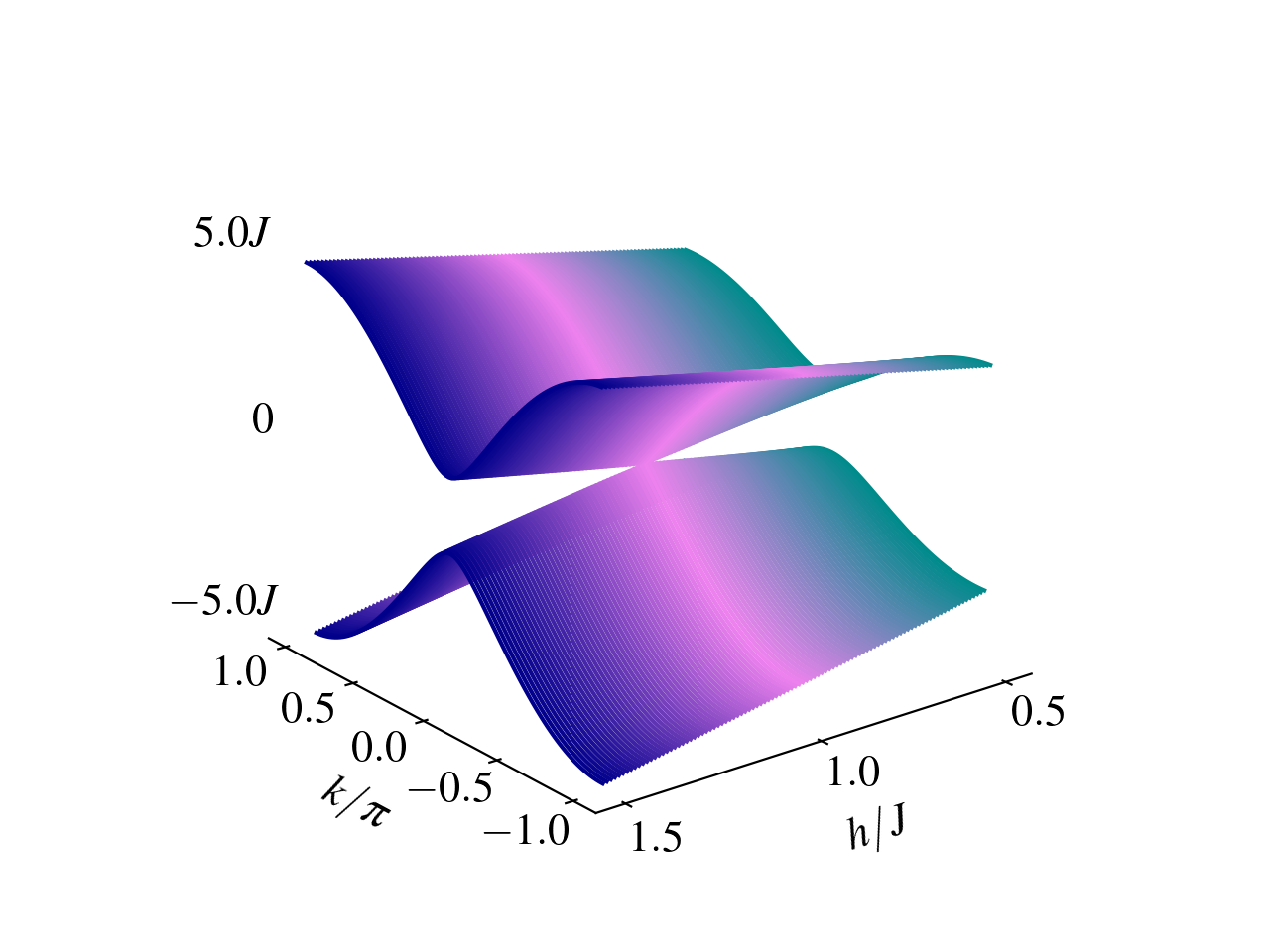}};		
		\node[anchor=south] (dispersion-TFIM) at ($(TFIM-contour.west)+(0,1cm)$)
		[font=\large, rotate=90]
		{$\epsilon_k$};

		\node[anchor=north west] (TFIM-contour-renormalized) at ($(dispersion-TFIM.south west)+(0cm,-3.5cm)$) {\includegraphics[width=0.9\columnwidth]{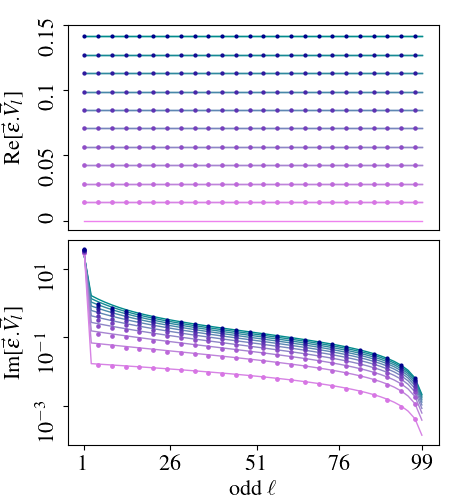}};	
    
    \end{tikzpicture}
    \caption{
        \textbf{Band structure and renormalized single-particle energies
        $\tilde{\epsilon}_{\ell}$ for an Ising chain under a transverse field with strength $h$.}
        The number of sites/levels is $L = 100$. Top contour shows the band dispersion as a
        function of $k$ and $h$. Bottom panels display the real and imaginary parts of the $\tilde{\epsilon}_{\ell}$'s for $\ell$ odd.
        All components with $\ell$ even vanish.
        The imaginary part decays as a function of $\ell$. Both real and imaginary parts are
        degenerate around $h_c = J$, i.e.
        $\tilde{\epsilon_\ell}[h-h_c<0] = \tilde{\epsilon_\ell}[h-h_c>0]$.
        At the critical point, the dominant renormalized energy is $\tilde{\epsilon}_1$, all
        other renormalized energies can be neglected in a first approximation. As a result, the
        MBDoS will be similar to the one of the 1D tight binding model.
    }
    \label{fig:new_single_particle_energies_TFIM}
\end{figure}

This can be observed in Fig. \ref{fig:new_single_particle_energies_TFIM}, panel (b), where we show the real and imaginary parts of $\tilde{\epsilon}_\ell$ for $h /J \in [0.5,1.5]$.

Using these previous results, we have all ingredients to compute the MBDoS for the applications aforementioned. In particular, for the 1D tight binding model and the transverse Ising chain at the critical point which have the same renormalized single body energies. As a result, they have the same MBDoS displayed on Fig. \ref{fig:MBDoS_TB1D}.

\section{Conclusion}
\label{sec:Conclusions}

In the present paper, we explored a novel approach to compute the many-body density of states
of quantum systems whose Hamiltonian can be mapped into non-interacting spinless fermions.

We show that the many-body spectrum can be expanded over the principal components of a filling
matrix encoding the allowed many body states. These principal components describe
 spectral properties of systems of spinless fermions only depending on the numbers of particles
and one body levels. This new spectral decomposition of many body states is weighted by a
renormalized single body band structure, which acts as a filter for relevant energy scales.

For gapless systems, such as the tight-binding and the critical transverse field Ising chains,
we demonstrated that only two renormalized energies are non zero. Even in more general
scenari, such as the square lattice or the Ising chain away from the critical point,
many renormalized energies still vanish. In all cases, this will significantly reduces the number
of relevant spectral components of the filling matrix involved in the calculation of the
many body density of states.

Our framework can be extended to include additional quantum numbers like spin, and to handle bosonic systems.

\section{Acknowledgements}

We acknowledge support from the Leverhulme Trust under grant RPG-2020-094.

\bibliography{MBDoS_refs.bib}

\end{document}